\documentclass[twocolumn,aps,pra,showpacs,superscriptaddress,nofootinbib]{revtex4-1}

\usepackage[latin1]{inputenc}
\usepackage{graphicx,epstopdf}
\usepackage{amsmath}
\usepackage{hyperref}
\usepackage{amsfonts}
\usepackage{amsthm}
\usepackage{amssymb}
\usepackage{color}
\usepackage{latexsym}
\usepackage{times,txfonts}

\newcommand{\fig}[1]{{Fig.}}

\begin{document}

\title{Non-Markovianity through quantum coherence in an all-optical setup}

%%%%%%%%%%%%%%%%%%%%%%%%%%%%%%%%%%%%%%%%%%%%%%%
%%%%%%%%%%%%%%---AUTHORS---%%%%%%%%%%%%%%%%%%%%
%%%%%%%%%%%%%%%%%%%%%%%%%%%%%%%%%%%%%%%%%%%%%%%
\author{M. H. M. Passos}
%\email{@if.edu.br }
\affiliation{
Instituto de Ci\^encias Exatas, Universidade Federal Fluminense, 27213-145 Volta Redonda, Rio de Janeiro, Brazil}

\author{P. C. Obando}
%\email{paolaconcha@if.uff.br}
\affiliation{Instituto de F\'isica, Universidade Federal Fluminense, Av. Gal. Milton Tavares de Souza s/n, Gragoat\'a, 24210-346, Niter\'oi, RJ, Brazil}

\author{W. F. Balthazar}
%\email{@if.edu.br }
\affiliation{
Instituto Federal do Rio de Janeiro, 27213-100 Volta Redonda, Rio de Janeiro, Brazil}

\author{F. M. Paula}
%\email{fagnerm@utfpr.edu.br }
\affiliation{N\'ucleo de Ci\^encias F\'{\i}sicas, Universidade Tecnol\'ogica Federal do Paran\'a, Rua Cristo rei 19, Vila Becker, 85902-490, Toledo, PR, Brazil}

\author{J. A. O. Huguenin}
%\email{jose$\_$huguenin@id.uff.br}
\affiliation{
Instituto de Ci\^encias Exatas, Universidade Federal Fluminense, 27213-145 Volta Redonda, Rio de Janeiro, Brazil}

\author{M. S. Sarandy}
%\email{msarandy@if.uff.br}
\affiliation{Instituto de F\'isica, Universidade Federal Fluminense, Av. Gal. Milton Tavares de Souza s/n, Gragoat\'a, 24210-346, Niter\'oi, RJ, Brazil}

\date{\today}

\begin{abstract}
We propose an all-optical experiment to quantify non-Markovianity in an open quantum system through quantum coherence of a single quantum bit. 
We use an amplitude damping channel implemented by an optical setup with an intense laser beam simulating a single-photon polarization. 
The optimization over initial states required to quantify non-Markovianity is analytically evaluated. 
The experimental results are in a very good agreement with the theoretical predictions.
\end{abstract}

\maketitle

%%%%%%%%%%%%%%%%%%%%%%%%%%%%%%%%%%%%%%%%
%% ---Introduction
%%%%%%%%%%%%%%%%%%%%%%%%%%%%%%%%%%%%%%%%
Quantum coherence is a fundamental feature of quantum mechanics and is an
important physical resource in quantum information \cite{Streltsov2016,Hu2017,baumgratz}. Quantum optical methods provide an important set of tools for the manipulation of coherence, and indeed, at its basis 
lies the formulation of the quantum theory of coherence \cite{glauber1963,sudarshan1963,Streltsov2016}.
Optical setups are largely used to investigate quantum information tasks once one can encode a quantum bit (qubit) in the degrees of 
freedom of light, such as propagation path, polarization, and transverse modes. For instance, entanglement \cite{ep1, Davichannels}, 
quantum computation \cite{swdeutsch,quantcomp}, quantum gates \cite{qgp1,qgp2}, 
quantum cryptography \cite{qcrypt}, and teleportation \cite{qtelep} have been investigated by encoding qubits in the degrees of 
freedom of light. In parallel, recent works have been exploring linear optical setup with an intense laser beam for investigate quantum 
features. This interesting approach has been used to investigate the emergence of topological phases in the evolution of a pair of entangled qubits \cite{topo}, environment induced entanglement \cite{environ},  Bell's  inequality \cite{bell1, bell3}, Mermin's inequality for entangled tripartite system \cite{tript}, cryptography \cite{ccrypt}, and conditional operations 
that emulate quantum gates \cite{cqg1, cqg3}. In this scenario, linear optical circuits associated to intense laser beams can be used 
to simulate single-photon experiments. %Such approach is a suitable tool to simulate test the quantum features that we deal with in this Letter.

Inspired by the recent developments about the quantitative characterization of coherence \cite{baumgratz}, our purpose 
here is to investigate through both theory and experiment the application of quantum coherence of a single qubit as a measure of non-Markovianity 
in an open quantum system. Classical and quantum correlations between two or more subsystems have been previously applied to characterize a non-Markovian evolution~\cite{paulapaola,Rivas:17,breuer103}. While such correlations characterize the quantum features of a system with at least two parties, quantum coherence is already defined for a single system \cite{Streltsov2016,Hu2017,baumgratz}, 
which provides the simplest scenario to access quantum superposition. By taking into account that the non-Markovian behavior can be characterized via quantum coherence \cite{chanda2016}, as an example, we analize a single-qubit  non-Markovian amplitude damping (AD) channel theoretically and experimentally. We experimentally realize this system through an optical setup with an 
intense laser beam simulating a single-photon polarization, with the system-environment interaction encoded in the propagation path. The experiment provides a convenient framework to illustrate the quantification of non-Markovianity through the revivals of the single-qubit coherence, which are controllable in terms of the non-Markovian strength.

%%%%%%%%%%%%%%%%%%%%%%%%%%%%%%%%%%%%%%%%
%% ---Quantum coherence measures
%%%%%%%%%%%%%%%%%%%%%%%%%%%%%%%%%%%%%%%%
Let us begin our discussion describing the incoherent states and the incoherent operations. Coherence is naturally a basis-dependent concept \cite{Streltsov2016}. 
For this reason, we need first  to define an orthonormal {\it local} reference basis  $\{\vert r\rangle\} = \{\vert r_1\rangle\ \otimes \cdots \otimes \vert r_N\rangle\}$ 
for an $N$-partite system represented in a $d$-dimensional Hilbert space $\mathcal{H}$. 
The density matrices acting on $\mathcal{H}$ that are diagonal in this specific basis form the set of incoherent density operators $\mathcal{I}$ 
acting on $\mathcal{H}$. Therefore, all density operators of the form
$\delta= \underset{r=1}{\sum^{d}}p_r\vert r\rangle\langle r\vert$, with the set $\{p_r\}$ denoting a probability distribution, 
are incoherent $(\delta\in\mathcal{I})$ \cite{baumgratz}. For Markovian quantum open systems, the quantum operations are described by completely positive and trace-preserving 
(CPTP) maps in terms of a set of Krauss operators $\{K_n \}$ satisfying 
$\sum_n K^{\dagger}_{n}K_n=\mathbb{I}$ \cite{nielsen}.  By definition, incoherent CPTP (ICPTP) operations are a subset of quantum operations with the restriction  $K_n\mathcal{I}K^{\dagger}_{n}\subset \mathcal{I}$ for all $n$. Thus, ICPTP operations, which act as $\Phi_{\mathrm{ICPTP}}(\rho) = \sum_nK_n\rho K^{\dagger}_{n}$, transform incoherent states into incoherent states, i.e., for any $\delta\in\mathcal{I}$, $\Phi_{\mathrm{ICPTP}}(\delta)\in\mathcal{I}$ \cite{baumgratz,Streltsov2016}. A measure of coherence $C$  is a nonnegative function which vanish for incoherent states and  is a nonincreasing monotone under incoherent ICPTP operations, $ C(\rho) \geq C(\Phi_{\mathrm{ICPTP}}(\rho))$. Some examples of $C(\rho)$ which fulfills these conditions are the relative entropy of coherence, $l_1$ norm of coherence \cite{baumgratz}, geometric coherence, coherence monotones from entanglement \cite{Streltsov2015}, among others\cite{Streltsov2016,napoli2016,shao2015fidelity}. An important class of coherence measures is given by a (pseudo-)distance between  $\rho$ and the closest incoherent  state $\delta_{min}$: $C(\rho)=\mathcal{D}(\rho,\delta_{min})$, where $\mathcal{D}$ is a contractible (pseudo-)distance measure \cite{baumgratz}. 
An example of a distance-based coherence is the trace norm of coherence, which for the one-qubit case is given by $C(\rho)=\parallel \rho - \delta_{min}\parallel_1 =  2\vert \rho_{12}\vert$,
where $\parallel A\parallel_1=\mathrm{tr}\sqrt{A^{\dagger}A}$ denotes the trace norm of the matrix $A$ and $\rho_{12}$ denoting the off-diagonal element of the one-qubit density matrix $\rho$ \cite{baumgratz,shao2015fidelity}.

Markovian evolution washes out correlations in a quantum system, making quantum correlation measures monotonic under local CPTP maps~\cite{breuerOQS,breuer103,breuercolloquium,luofusong,paulapaola}. It has been proved that basis-independent measures of quantum coherence may be exactly equivalent to entanglement \cite{Streltsov2015} and other quantum correlations \cite{xi, yao}. 
Following the same line of thought, based on the monotonically decreasing behavior of  quantum coherence measures under ICPTP maps, we have that for Markovian dynamics, it follows that $d C(\rho(t))/dt\leq 0$, where $C(\rho)$ is a proper 
quantum coherence measure. Thus, any violation of this monotonicity  $d C(\rho(t))/dt> 0$ at any time $t$ will provide an indication 
of non-Markovianity for an arbitrary quantum coherence measure defined in terms of any non-diagonal reference basis. From this non-monotonicity of quantum coherence measures, we can define a quantifier of non-Markovianity as
\begin{equation}\label{Eq:NCoherence}
N_C(\Phi)=\underset{\rho(0)}{\mathrm{max}}\underset{\frac{dC(\rho(t))}{dt}>0}{\int}\frac{d}{dt}C(\rho(t)) dt, 
\end{equation} 
where the maximization is taken over all initial states $\rho(0)$ \cite{chanda2016}. Hence, $N_C(\Phi)$ quantifies the degree of non-Markovianity for dynamical maps that preserve incoherence. It leads to the interpretation of the reservoir memory effect as a backflow of quantum coherence on the initial state, after the state has been subject to a noisy channel for a certain time.  

In order to quantitatively analyze of $N_C(\Phi)$, we will consider the coherence of a single qubit in the computational basis $\{\vert 0 \rangle, \vert 1 \rangle \}$ under a non-Markovian AD-channel, whose dynamics is given by the damped Jaynes-Cummings model on resonance, which is often used to describe a two-level atom interacting with a single cavity mode coupling to a bosonic reservoir~\cite{breuerOQS,bellomo2007non}. The incoherent Kraus operators of this channel are defined by: 
$K_0= \vert 0\rangle\langle 0\vert + \sqrt{p(t)} \vert 1\rangle\langle 1\vert,\: K_1=\sqrt{1-p(t)}\vert 0\rangle\langle 1\vert$, where $p(t)= e^{-\Gamma t}\left\lbrace  \cos\left(\frac{\Gamma\sqrt{\alpha-1}}{2}t\right)+ \frac{1}{\sqrt{\alpha-1}}\sin\left(\frac{\Gamma\sqrt{\alpha-1}}{2}t\right) \right\rbrace^2 $.
Here $\alpha =2\gamma/\Gamma$, being  $\gamma$ the system-reservoir coupling constant and $\Gamma$ the decay rate of the qubit \cite{bellomo2007non}. In the weak-coupling regime, i.e. for  $0\leq\alpha \leq 1$, $p(t)$  is monotonically decreasing (it is essentially an exponential decay controlled by $\gamma$). On the other hand, in the 
strong-coupling regime $\alpha > 1$, $p(t)$ exhibits a non-monotonic behavior and the non-Markovian effects become relevant. 

\begin{figure}[ht!]
 \begin{center}
\includegraphics[height=4.5cm,width=7cm]{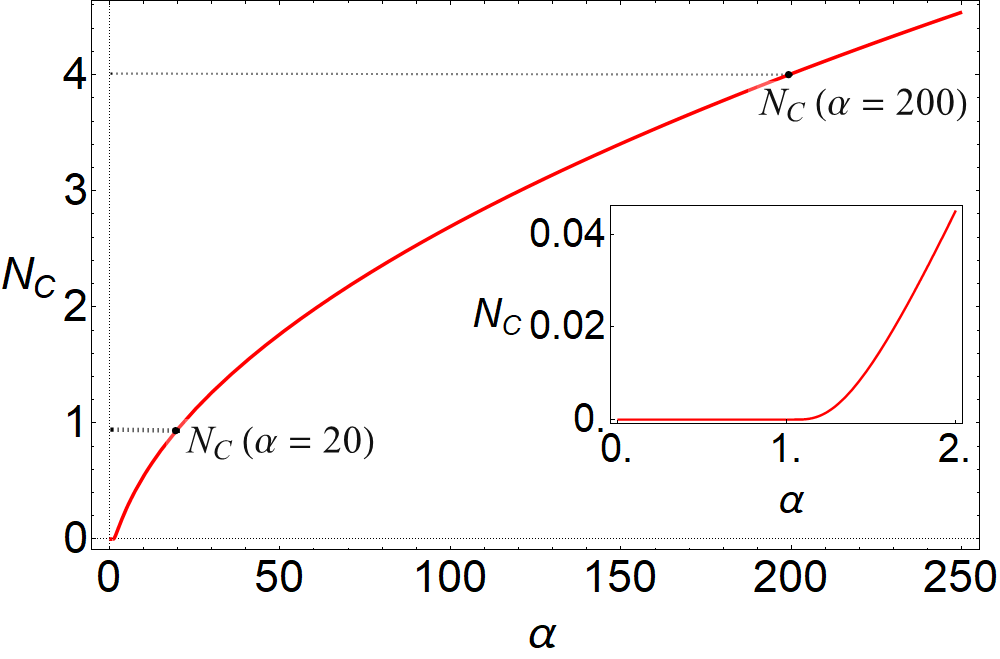}
 \end{center}
\caption{Degree of non-Markovianity $N_C$ as a function of  the parameter $\alpha$ for an one-qubit system  under a non-Markovian AD noise. 
The values of $N_C$ for $\alpha=20$ and $\alpha=200$ will be particularly analyzed in the experimental realization.
In the inset, we detail the evolution of $N_C$ for the regime $0\leq \alpha\leq 2$. }
  \label{Fig:Nc}
\end{figure}

Let us now calculate $N_C(\Phi)$ for the non-Markovian AD-channel. In particular, the trace norm of coherence for this channel is given by $C(t)= C(0)\sqrt{p(t)}$,  where $C(0)=2\vert \rho_{12}(0)\vert$. For $0\leq \alpha \leq 1$, $C(t)$ is monotonically decreasing and, consequently, $N_C(\alpha)=0$. On the other hand, for $\alpha >1$, $C(t)$ shows an oscillatory decay with the maximums and minimums occurring at  $t_{m}^{max}= 2\pi\Gamma^{-1}\left(\alpha-1\right)^{-1/2}m$ and  $t_{m}^{min}= t_{m+1}^{max}-2\Gamma^{-1}\left(\alpha-1\right)^{-1/2} \mathrm{arctan}\sqrt{\alpha-1}$ $(m= 0,1,2,..)$, respectively. In this case, Eq.~(\ref{Eq:NCoherence}) can be written as
\small
\begin{equation}
N_C(\alpha)=\underset{\rho(0)}{\mathrm{max}}\sum^{\infty}_{m=1}\left[ C(t_m^{max})-C(t_{m-1}^{min}) \right]= \underset{\rho(0)}{\mathrm{max}}C(0)\sum^{\infty}_{m=1}e^{-\pi m/\sqrt{\alpha-1}}
\end{equation} 
\normalsize
being $\lbrace(t_m^{min},t_{m-1}^{max})\rbrace$ $(m\neq 0)$ the set of all intervals of time such that $dC(\rho)/dt>0$. The maximization over $\rho(0)$ is satisfied when $C(0)=1$, i.e., when $|\rho_{12}(0)|=1/2$. In addition, $\sum^{\infty}_{m=1} e^{-\pi m/\sqrt{\alpha-1}}= \left(e^{\pi/\sqrt{\alpha-1}}-1\right)^{-1}$. Thus, we find a compact analytical expression for the degree of non-Markovianity in terms of the parameter $\alpha$, reading
\small
\begin{equation}
N_C (\alpha) = \left\{ \begin{matrix} \left(e^{\frac{\pi}{\sqrt{\alpha-1}}}-1\right)^{-1} & \alpha > 1,
\\ 0 &  0 \leq \alpha \leq 1. \end{matrix}\right.
\label{Nc17}
 \end{equation}
\normalsize
The behavior of $N_C(\alpha)$ is illustrated in Fig. \ref{Fig:Nc}. In particular, $N_C(\alpha) \approx \sqrt{\alpha}/\pi$ in the strong non-Markovian regime ($\alpha >> 1$).
%%%%%%%%%%%%%%%%%%%%%%%%%%%%%%%%%Figure%%%%%%%%%%%%%
%%%%%%%%%%%%%%%%%%%%%%%%%%%%%%%%%%%%%%%%%%%%%%%%%%%%%%%%%%%%%%
%%%%%%%%%%%%%%%%%%%% STATES MAP %%%%%%%%%%%%%%%%%%%%%%%%%%%%%%
%
The effect of the AD-channel on a single qubit system in the maximally coherent state $|\psi_+ \rangle = \frac{1}{\sqrt{2}}[|0\rangle_S + |1\rangle_S]$  can be regarded through the following map
\begin{small}
\begin{equation}
 \label{supmap}   
|\psi_+ \rangle_S \; |0 \rangle_R  \rightarrow
\frac{1}{\sqrt{2}} [\;|0\rangle_S \; |0 \rangle_R  + \sqrt{p(t)} \;|1 \rangle_S \; |0 \rangle_R  
+ \; \sqrt{1-p(t)} |0 \rangle_S \; |1 \rangle_R\;].
\end{equation}
\end{small}
where the label $S$ and $R$ stands for system and reservoir states, respectively. 
Looking at the right-hand side of Eq.~(\ref{supmap}), we can see that the ground state of the qubit ($|0\rangle_S$) remains unchanged along the evolution (first term). On the other hand, the excited state of the qubit ($|1\rangle_S$) can decay with probability $[1-p(t)]$. Note that for $t \rightarrow \infty$, $[1-p(t)] \rightarrow 1$, which means the excited qubit will decay after long time interaction with the channel. For this case, the superposition is lost and the system state decays to its ground state, for which coherence vanishes. As discussed before, depending on the system-reservoir coupling constant $\alpha$ we can achieve Markovian or Non-Markovian behavior. The Non-Markovian behavior can be witnessed by observation of oscillations on the coherence values along the evolution.

In order to experimentally investigate Non-Markovian signatures by means quantum coherence in the AD channel, we used propagation direction (path) and the polarization degrees of freedom of light. The system states are encoded in the polarization and the reservoir is encoded in the path. Regarding the state of polarization of a single-photon as a two-level system, we can associate the horizontal and vertical polarization state with the ground state ($ | H \rangle \equiv | 0 \rangle_S$) and excited state ($ | V \rangle \equiv | 1 \rangle_S$), respectively. Regarding the reservoir, the ground state $|0 \rangle_R$ is encoded in the output of the AD-channel circuit that does not change the input state of the system (polarization).
% can associate orthogonal directions, $\vec{k_0}$ and $\vec{k_1}$, corresponding to the reservoir ground state $|0 \rangle_R$ and excited state ($|1 \rangle_R$), respectively.
%By using the polarization labels for the system states, the maps given by Eq. (\ref{Eq:map01}) and (\ref{Eq:map02}) can be regarded as follows 
%$|H \rangle_S \; |0 \rangle_R \rightarrow |H \rangle_S \; |0 \rangle_R , $
%

%$|V \rangle_S \; |0 \rangle_R \rightarrow \sqrt{p(t)}\;|V \rangle_S \; |0 \rangle_R + \sqrt{1-p(t)}\; |H \rangle_S \; |1 \rangle_R .$

%Then, for a maximally coherent superposition state ($|\psi_+\rangle$) the map presented in Eq. (\ref{Eq:mapsup}) can be written as  $|\psi_+ \rangle_S \; |0 \rangle_R  \rightarrow \frac{1}{\sqrt{2}} [\; |H\rangle_S \; |0 \rangle_R + \sqrt{p(t)}\;|V \rangle_S \; |0 \rangle_R + \sqrt{1-p(t)}\; |H \rangle_S \; |1 \rangle_R \;]$.

\begin{figure}[htp!]
 \begin{center}
 \includegraphics[scale=0.34,clip,trim=0mm 47mm 0mm 50mm]{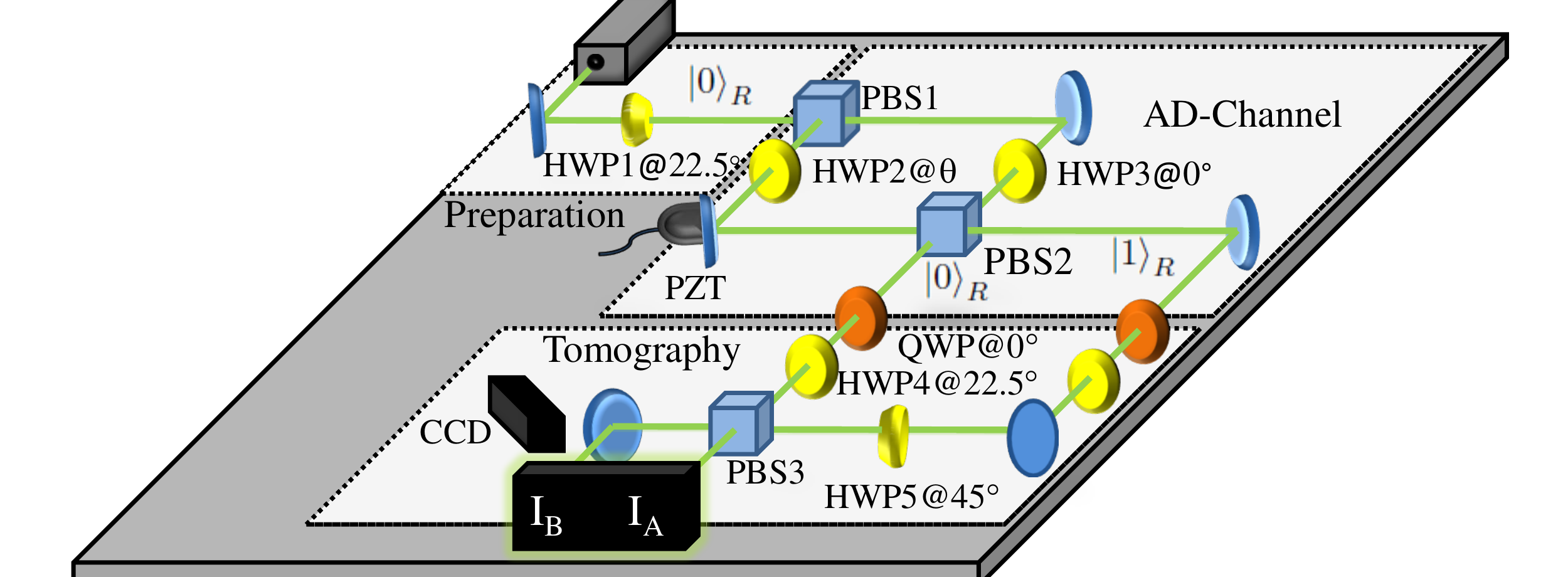} % SEUTUP
\end{center}
\caption{Experimental setup.}
 \label{fig:setup}
\end{figure}

We performed the experiment with an intense laser beam once it simulates the single-photon experiment and its results present the essence of the phenomenon we are studying. The use of intense laser beam to simulate quantum tasks has been used frequently in the literature \cite{topo, environ, bell1, bell3, tript, ccrypt, cqg1, cqg3}. In this approach, a laser beam polarized at $+45^\circ$ can be regarded as the analog of the maximally coherent state, $|\psi_+ \rangle_S$. Then, we represent the H and V polarization respectively as $|H\rangle$ and $|V\rangle$, in order to directly associate the polarization quantum state of a single photon.
The linear optical circuit used in the experiment is presented in Fig.~\ref{fig:setup}. A DPSS laser ($532 nm, 1.5 mw$ power, H-polarized) passes through a half wave plate (HWP1@22.5) with its fast axis making an angle of $22.5^\circ$ by respect the horizontal in order to prepare a $+45^\circ$ polarized beam, i.e., the analogue to the state $|\psi_+ \rangle_S \equiv |+\rangle_S = \frac{1}{\sqrt{2}} \left(\;|H\rangle+|V\rangle \; \right)$.
The resulting path is associated with the reservoir ground state. Then, after the HWP1, we have the input state $|+\rangle_S |0\rangle_R$. This step is highlighted  in Fig.~\ref{fig:setup} in the block "Preparation". It is worth to mention that we can prepare different input states by rotating HWP1.

\begin{figure}[htp!]
 \begin{center}
 \includegraphics[scale=0.45,clip,trim=45mm 72mm 10mm 20mm]{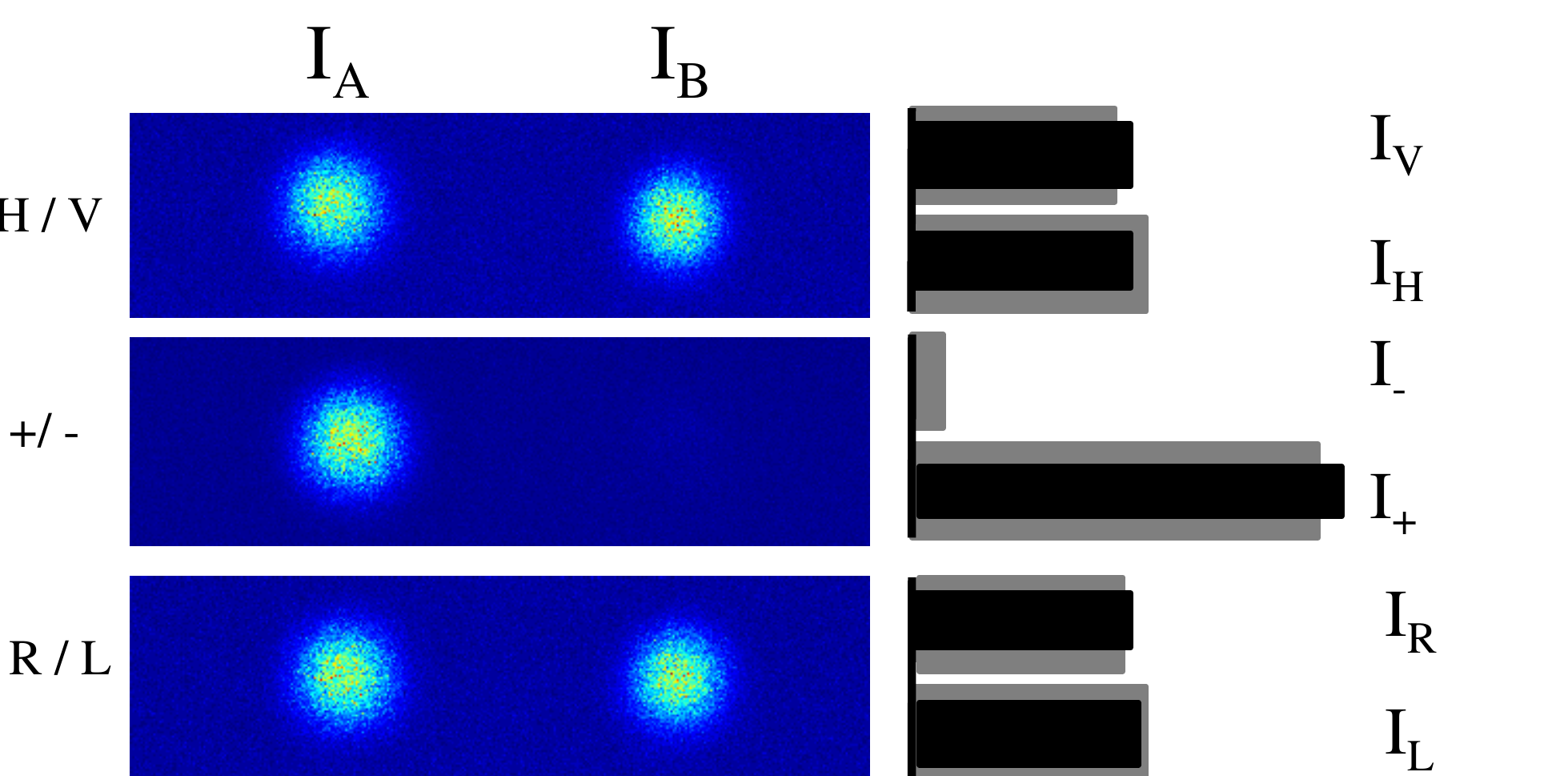} % TYPICAL IMAGES
 \end{center}
\caption{Image false color for tomography of the initial state $|+\rangle_S$. Normalized intensity of basis components are represented by black (theory) and grey (experiment) bars.}
  \label{fig:images}
\end{figure}

The AD-channel starts with a polarized beam splitter (PBS1) that transmits H-polarization ($|H\rangle$) and reflects the V-polarization($|V\rangle$). The reflected  component passes through the HWP2@$\theta$, (with $\theta$ measured by respect the vertical). By setting $\theta=0^\circ$, polarization is not changed and the beam is reflected in PBS2 in the path $|0\rangle_R$ (showed in Fig.\ref{fig:setup}). In the transmitted arm of PBS1 a fixed HWP3@$0^\circ$ maintain polarization unchanged and this component also leave the channel in path $|0\rangle_R$.
A piezoelectric ceramic (PZT) placed in one mirror controls the difference of phase $\Delta \phi$ between the two arms. This device allows us to perform a coherent superposition of $ |H\rangle$ and $ |V\rangle$  components in the output path $|0\rangle_R$. For $\Delta \phi = 0$ we recover the $+45^\circ$ polarized light, i.e., $|+\rangle_S |0\rangle_R$ . This case simulates the instant $t=0$, when the system has not yet interacted with the reservoir and the coherence is kept maximal.
The HWP2@$\theta$ emulates the parameter $p(t)$. After HWP2@$\theta$ we have the following transformation: $|V\rangle \rightarrow \sin(2\theta)\; |H\rangle + \cos(2\theta)\; |V\rangle.$
Then, by adjusting $\Delta \phi = 0$, the polarization of the beam in the path $|0\rangle_R$ is $|\phi_0(\theta)\rangle_{S}\;|0\rangle_R = \frac{1}{\sqrt{2}}\left[\; |H\rangle + \cos(2\theta)\; |V\rangle \; \right]\;|0\rangle_R.$
On the other hand, for the output path associated to $|1\rangle_R$ we have only the horizontal component of the polarization produced by HWP2@$\theta$ $|\phi_1(\theta)\rangle_{S}\;|1\rangle_R = \frac{1}{\sqrt{2}} \left[ \;\sin(2\theta)\; |H\rangle \; \right]\;|1\rangle_R$,
which corresponds to the damping produced by the channel with the reservoir state receiving a quantum of energy. The complete map of the AD-channel circuit can be written as
\begin{eqnarray}
\label{Eq:circuitmap}
|\psi_+ \rangle_S \; |0 \rangle_R   & \rightarrow & |\phi_0(\theta)\rangle_{S}\; |0\rangle_{R} + |\phi_1(\theta)\rangle_{S}\; |1\rangle_{R} \nonumber \\ 
 & \rightarrow &
\frac{1}{\sqrt{2}} [\; |H\rangle_S \; |0 \rangle_R + \cos(2\theta)\;|V \rangle_S \; |0 \rangle_R \\ \nonumber
& + & \sin(2\theta)\; |H \rangle_S \; |1 \rangle_R \;].
\end{eqnarray}
By comparing Eq. (\ref{Eq:circuitmap}) and Eq. (\ref{supmap}) we identify $\sin(2\theta) = \sqrt{1-p(t)}$. Then, for each $\gamma t$ we can associate a corresponding angle $\theta$ of the HWP2@$\theta$. Note that $p(t)$ is emulated by the optical setup.
The states associated with the paths $|0\rangle_R$ and $|1\rangle_R$ of the AD-channel output are directed to the "tomography" block that performs a polarization state tomography~\cite{tomography} for the case where we trace-out the environment. PBS3 measure the polarization in $\{|H\rangle, |V\rangle\}$ basis, HWP4@$22.5^\circ$ associated with PBS3 proceed measurements in the diagonal basis ( $\{|+\rangle, |-\rangle\}$), and the sequence of a quarter wave plate QWP@$0^\circ$,  HWP4@$22.5^\circ$, and PBS3 measure in the  right-handed and left-handed circular polarization basis ($\{| R\rangle, |L\rangle\}$). HWP5@$45^\circ$ is used to make $|H\rangle_S$ be reflected by PBS3. The intensity of each component is projected on a screen and recorded in a single image by a charged-coupled-device (CCD) camera. The normalized intensity $I_\beta/I_T$ ($\beta= A, B$, $T \equiv$Total) plays the role of the probabilities in the density matrix $\rho_{ij}$ reconstruction. Depending on the measurement basis $A=H, +,$ and $R$, while $B=V, -,$ and $L$, respectively. Fig.~\ref{fig:images} shows the resulting intensities for $\theta = 0^\circ$, which corresponds to the initial state in AD-channel ($t=0$). The tomographic measured basis are indicated to the left of each image. Bars at the right of each the image illustrate the normalized intensity,
that was used to obtain the corresponding density matrix \cite{tomography} and calculate the coherence as $C(t)=2|\rho_{12}(t)|$. For the initial state we obtained $C(0)=0.98 \pm 0.03$, very close to the expected unit value. The error comes from the limited visibility of the interferometer and the intensity sensitivity of the CCD camera.

\begin{figure}[htp!] 
 \begin{center}
\includegraphics[scale=0.63,clip,trim=55mm 55mm 0mm 51mm]{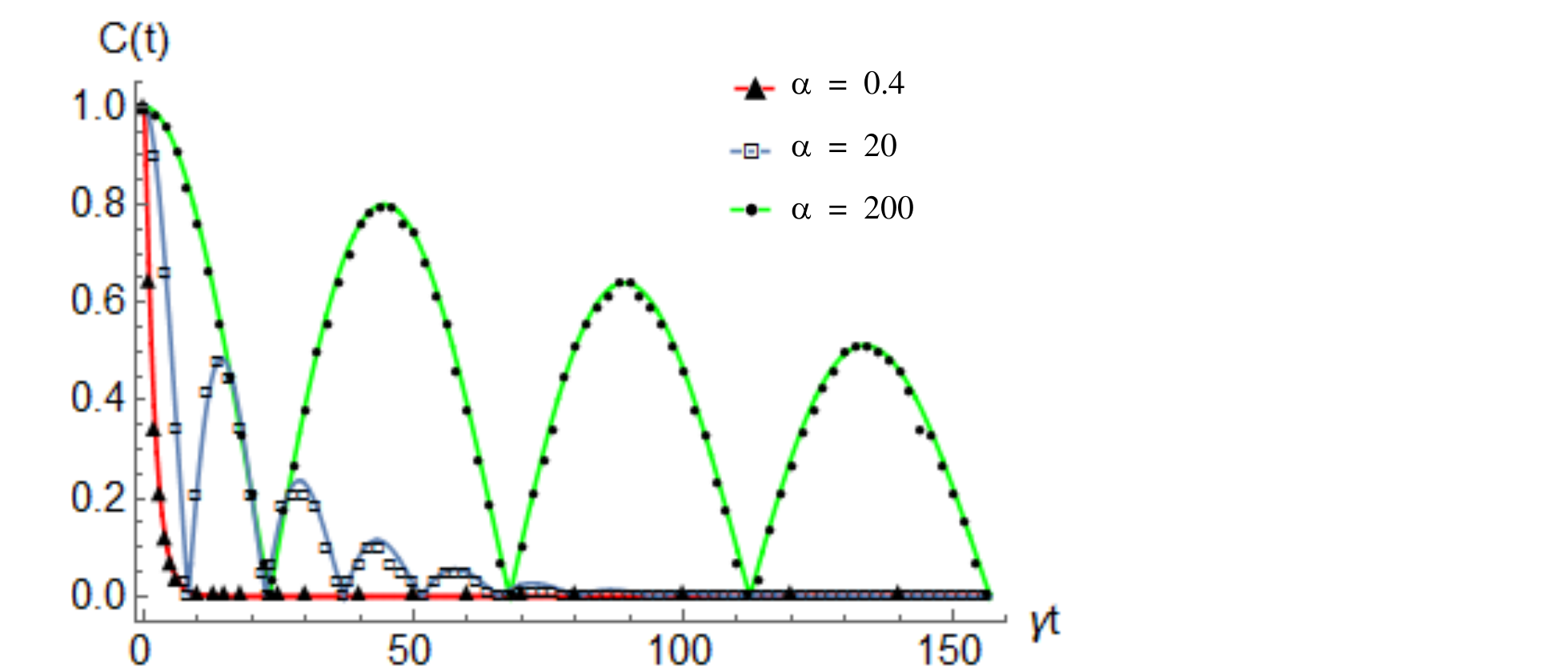} % GRAPHIC
 \end{center}
\caption{Experimental coherence for Markovian behavior (black triangles), weak non-Markovian behavior (white squares), and strong non-Markovian behavior (black circles). The solid lines represent the theoretical curves, for $\alpha=0.4$ (red),  $\alpha=20$ (blue), and  $\alpha=200$ (green). }
  \label{fig:graphic}
\end{figure}

Now we are able to evaluate non-Markovian signatures in the AD channel by means of coherence. It is worth to mention that $p(t)$ has different behavior for Markovian and Non-Markovian evolution. One can achieve both evolutions by controlling the coupling constants $\Gamma$ and $\gamma$ in $p(t)$ and we can find the correspondent $\theta$ for each $\gamma t$. By performing the tomography of the output states, we reconstruct the corresponding density matrix and calculate the coherence % according Eq. (\ref{Eq:coherencequbit}).
for each $\gamma t$. The results are shown in Fig.~\ref{fig:graphic}, with the discrete points representing 
the experimental data and the solid curves their corresponding theoretical counterparts.The agreement is remarkable. 

By setting $\Gamma = 5\gamma$, we found $\alpha=0.4$, which implies in Markovian behavior, as indicated by Eq. (\ref{Nc17}). As it can be seen from the plot, 
coherence vanishes fast, presenting an exponential decay as expected for Markovian evolution. Let us verify the results for an increment of the coupling. 
By setting $\Gamma = 0.1\gamma$,  we obtain $\alpha = 20$, leading to small oscillations of the coherence before totally vanishing, in agreement with a smooth non-Markovian behavior expected for this value of $\alpha$ according to Eq. (\ref{Nc17}).
By setting $\Gamma = 0.01\gamma$, we have $\alpha=200$. Then, according to Eq. (\ref{Nc17}), a strong Non-Markovian evolution is expected to occur. This is indeed observed, with the presence of strong revivals of coherence.  

Let us now discuss the quantifier of non-Markovianity $N_C(\alpha)$, evaluated from the experimental data. The values of $\alpha$ considered in the experiment are pointed out in the theoretical plot exhibited in Fig~\ref{Fig:Nc}. For the Markovian evolution ($\alpha=0.4$) no revivals are observed, and $N_C(\alpha) = 0$, as expected by Eq. (\ref{Nc17}). For the weak non-Markovian behavior ($\alpha = 20$) we can clearly identify five local maximum revivals of coherence in Fig.\ref{fig:graphic}. For this case we experimentally found $N_C(\alpha) = 0.86$, which is close to the value $N_C(\alpha) = 0.92$ obtained from the theoretical expression given by Eq. (\ref{Nc17}) when 5 revivals are considered. On the other hand, by integrating from 0 to $ t\rightarrow \infty$ (infinite revivals) we obtain $N_C(\alpha) = 0.95$. This difference between the total integration and the finite revivals become more accentuated for larger $\alpha$. For the case when $\alpha=200$, we experimentally observed 3 revivals for the time range of the experiment, which produces $N_C(\alpha) = 1.90$ in agreement with the theoretically calculated value $N_C(\alpha) = 1.95$ for 3 revivals. However, in this case, the comparison between calculation using experimental values and integration of Eq. (\ref{Nc17}) is not appropriated, since only 3 revivals have been considered in the experiment. By integrating Eq. (\ref{Nc17}) we find $N_C(\alpha) = 4.01$ [see also Fig~\ref{Fig:Nc}]. Hence, for a strong non-Markovian regime, the computation of $N_C$ requires that the observation time is sufficiently large in order to the experimental values for finite revivals to reproduce the theoretical result predicted by Eq. (\ref{Nc17}) for infinite revivals.

%%%%%%%%%%%%%%%%%%%%%%%%%%%%%%%%%%%%%%%%
% --- CONCLUSIONS ---
%%%%%%%%%%%%%%%%%%%%%%%%%%%%%%%%%%%%%%%%

%\section{ Conclusions}\label{section:conclutions}

We have presented an all-optical experiment to observe non-Markovian behavior through a coherence-based quantifier of non-Markovianity N$_C$($\alpha$). We have shown that, under the allowed incoherent operation criterion, the monotonicity of the valid coherence measure may be affected by a partial backflow of the previously lost information of the system to the environment. An advantage of this method is that, instead of using ancillary subsystems, it involves only a single qubit, implying a simpler process of optimization and empirical realization. It is worth emphasizing that the coherence measure of non-Markovianity is applicable beyond the AD channel, holding for any incoherence-preserving channel. 

%%%%%%%%%%%%%%%%%%%%%%%%%%%%%%%%%%%%%%%%
% ---Funding Information ---
%%%%%%%%%%%%%%%%%%%%%%%%%%%%%%%%%%%%%%%%

\textbf{Funding}. Conselho Nacional de Desenvolvimento Cient\'{\i}fico e Tecnol\'ogico (CNPq); 
Coordena\c{c}\~ao de Aperfei\c{c}oamento de Pessoal de N\'{\i}vel Superior (CAPES);
Funda\c{c}\~ao Carlos Chagas Filho de Amparo \`a Pesquisa do Estado do Rio de Janeiro (FAPERJ); and the Brazilian National Institute for Science and Technology of Quantum Information (INCT-IQ).


\begin{thebibliography}{1}

%Quantum coherence%%%%%%%%%%%%%%%%%%%%%%%%%%%%%%%

% Colloquium: Quantum coherence as a resource
\bibitem{Streltsov2016} A. Streltsov, G.  Adesso, M. B. Plenio, Rev. Mod. Phys. \textbf{89}, 041003 (2017). %Review

% Quantum coherence and geometric quantum discord
\bibitem{Hu2017} M. L. Hu, X.  Hu, J. C.  Wang, Y.  Peng, Y. R. Zhang, H. Fan, Phys. Rep. 762, 1-100 (2018). %Review

% Quantifying Coherence
\bibitem{baumgratz}	T. Baumgratz, M. Cramer, M. Plenio, Phys. Rev. Lett. \textbf{113}, 140401 (2014).


% Coherent and Incoherent States of the Radiation Field
\bibitem{glauber1963} R. J. Glauber, Phys. Rev. \textbf{131},  2766 (1963). 

% Equivalence of Semiclassical and Quantum Mechanical Descriptions of Statistical Light Beams
\bibitem{sudarshan1963}	E. Sudarshan, Phys. Rev. Lett. \textbf{ 10}, 277 (1963). 

% Entanglement and conservation of orbital angular momentum in spontaneous parametric down-conversion
\bibitem{ep1}  S. P. Walborn, C. H. Monken, A. N. de Oliveira, R. S. Thebaldi, Phys. Rev. A \textbf{69}, 23811 (2004).

% Experimental investigation of the dynamics of entanglement: Sudden death, complementarity, and continuous monitoring of the environment
 \bibitem{Davichannels} Salles, A. and de Melo, F. and Almeida, M. P. and Hor-Meyll, M. and Walborn, S. P. and Souto Ribeiro, P. H. and Davidovich, L., Phys. Rev. A \textbf{78}, 2, 022322 (2008).

% Implementing the Deutsch algorithm with polarization and transverse spatial modes
 \bibitem{swdeutsch} S. P. Walborn, A. N. de Oliveira, C. H. Monken, J. Opt. B \textbf{7}, 288 (2005).

% Deterministic quantum computation with one photonic qubit
\bibitem{quantcomp} M. Hor-Meyll, D. S. Tasca, S. P. Walborn,  P. H. S. Ribeiro, M. M. Santos, E. I. Duzzioni, Phys. Rev. A \textbf{ 92}, 012337 (2015).

% Single-photon logic gates using minimal resources
\bibitem{qgp1} L. Qing, H. Bing, Phys. Rev. A \textbf{80}, 042310 (2009).

% High-Dimensional Single-Photon Quantum Gates: Concepts and Experiments
 \bibitem{qgp2}  A. Babazadeh, M. Erhard, F. Wang, M. Malik, R. Nouroozi, M. Krenn, A. Zeilinger, Phys. Rev. Lett. \textbf{119}, 180510, (2017). 

% Complete experimental toolbox for alignment-free quantum communication
 \bibitem{qcrypt} V. D'Ambrosio, E. Nagali, S. P. Walborn, L. Aolita, S. Slussarenko, L. Marrucci, F. Sciarrino, Nat. Commun. \textbf{3}, 961 (2012).

% Remote Preparation of Single-Photon “Hybrid” Entangled and Vector-Polarization States
 \bibitem{qtelep} J. T. Barreiro, T. C. Wei, P. G. Kwiat, Phys. Rev. Lett. \textbf{105}, 030407 (2010).

% Quantum teleportation in the spin-orbit variables of photon pairs
%\bibitem{qtelp2} A. Z. Khoury, P. Milman, Phys. Rev. A \textbf{83}, 060301 (2011).

% Topological phase for spin-orbit transformations on a laser beam
\bibitem{topo} C. E. R. Souza, J. A. O. Huguenin, P.Milman, and A. Z. Khoury, Phys. Rev. Lett. \textbf{99}, 160401 (2007)

% Topological phase structure of vector vortex beams
%\bibitem{topophase} C. E. R. Souza, J. A. O. Huguenin, A. Z. Khoury, J. Opt. Soc. Am. A \textbf{31}, 1007 (2014). 

% Experimental investigation of environment-induced entanglement using an all-optical setup
\bibitem{environ} M. H. M. Passos, W. F. Balthazar, A. Z. Khoury, M. Hor-Meyll, L. Davidovich, J. A. O. Huguenin, Phys. Rev. A \textbf{97}, 022321 (2018).

% Bell-like inequality for the spin-orbit separability of a laser beam
\bibitem{bell1} C. V. S. Borges, M. Hor-Meyll, J. A. O. Huguenin, A. Z. Khoury, Phys. Rev. A \textbf{82}, 033833 (2010).

%\bibitem{bell2} K. H. Kagalwala, G. Di Giuseppe, A. F. Abouraddy, B. E. A. Saleh, Nat. Photon. \textbf{7}, 72 (2012).

% Shifting the quantum-classical boundary: theory and experiment for statistically classical optical fields
\bibitem{bell3} X.-F. Qian, B. Little, J. C. Howell, J. H. Eberly, Optica \textbf{2}, 611 (2015).

% Tripartite nonseparability in classical optics
\bibitem{tript} W. F. Balthazar, C. E. R. Souza, D. P. Caetano, E. F. Galv\~ao, J.A. O. Huguenin, A. Z. Khoury, Opt. Lett. \textbf{41}, 5797 (2016).

% Quantum key distribution without a shared reference frame
\bibitem{ccrypt} C. E. R. Souza, C. V. S. Borges, A. Z. Khoury, J. A. O.Huguenin, L. Aolita, S. P. Walborn, Phys. Rev. A \textbf{77}, 032345 (2008).

% Spin–orbit laser mode transfer via a classical analogue of quantum teleportation
%\bibitem{ctelep} B. P. da Silva, M. A. Leal, C. E. R. Souza, E. F. Galv\~ao, A.Z. Khoury, J. Phys. B \textbf{49}, 055501 (2016).

% A Michelson controlled-not gate with a single-lens astigmatic mode converter
\bibitem{cqg1} C. E. R. Souza, A. Z. Khoury, Opt. Express \textbf{18}, 9207 (2010).

%\bibitem{cqg2} W. F. Balthazar,  D. P. Caetano, C. E. R. Souza,  J. A. O.Huguenin, Braz. J. Phys. \textbf{44}, 658 (2014).

% Conditional operation using three degrees of freedom of a laser beam for application in quantum information
\bibitem{cqg3}  W. F. Balthazar, J. A. O.Huguenin, J. Opt. Society America B \textbf{33}, 1649 (2016).


%Characterizing non-Markovianity%%%%%%%%%%%%%%%%%%%%%%%%%%%%%%%%%%%%%%%%

%\bibitem{rivashuelgaplennio2010} A. Rivas, S. F.  Huelga, M. B.  Plenio, Phys. Rev. Lett.  \textbf{105}, 050403 (2010). 

% Measure for the Degree of Non-Markovian Behavior of Quantum Processes in Open Systems
\bibitem{breuer103}	H.-P. Breuer, E.-M. Laine, J.  Piilo, Phys. Rev. Lett. \textbf{103}, 210401 (2009).

% Refined weak-coupling limit: Coherence, entanglement, and non-Markovianity
\bibitem{Rivas:17} A. Rivas, Phys. Rev. A \textbf{95}, 042104 (2017).

%\bibitem{fanchini2014non}	F. F. Fanchini, G.  Karpat, B. \c{C}akmak, L.  Castelano, G.  Aguilar, O. J.  Farias, S.  Walborn, P. S. Ribeiro, M. C. de Oliveira, Phys.  Rev.  Lett.  \textbf{112}, 210402 (2014).

% Non-Markovianity through multipartite correlation measures
\bibitem{paulapaola} F. M. Paula, P. C.  Obando, M. S.  Sarandy, Phys.  Rev.  A \textbf{93}, 042337 (2016).

% Delineating incoherent non-Markovian dynamics using quantum coherence
\bibitem{chanda2016} T. Chanda, S. Bhattacharya, Ann. Phys. \textbf{366}, 1 (2016).

% Quantum computation and Quantum Information
\bibitem{nielsen} M. A. Nielsen, I.  Chuang,  \textit{Quantum computation and Quantum Information}, Cambridge University Press, Cambridge, UK (2000).

\bibitem{Streltsov2015}	A. Streltsov, U.  Singh, H. S. Dhar, M. N Bera and G. Adesso, Phys. Rev. Lett. \textbf{115}, 020403 (2015). 

\bibitem{napoli2016} C. Napoli, T.  Bromley, M.  Cianciaruso, M.  Piani, N. Johnston, G.  Adesso, Phys. Rev. Lett. \textbf{116}, 150502 (2016).
%%%%%%%%%%%%%%%%%

% Fidelity and trace-norm distances for quantifying coherence
\bibitem{shao2015fidelity}	L.-H. Shao, Z. Xi, H.  Fan, Y.  Li, Phys. Rev. A \textbf{91}, 042120 (2015)

%%%%%%%%%%


% Colloquium: Non-Markovian dynamics in open quantum systems 
\bibitem{breuercolloquium} H.-P. Breuer, E.-M. Laine, J. Piilo, B. Vacchini, Rev. Mod. Phys.  \textbf{88}, 021002 (2016).

% Quantifying non-Markovianity via correlations
\bibitem{luofusong} S. Luo, S.  Fu, H. Song, Phys. Rev. A \textbf{86}, 044101 (2012).


% The theory of open quantum systems
\bibitem{breuerOQS} H.-P Breuer, and F. Petruccione. \textit{The theory of open quantum systems}. Oxford University Press on Demand, (2002).



%Quantum coherence and correlations in quantum system.
\bibitem{xi} Xi, Zhengjun, Yongming Li, and Heng Fan. Scientific reports 5, 10922 (2015)

%Quantum coherence in multipartite systems
\bibitem{yao} Y. Yao, X. Xiao, L. Ge, C. Sun, Phys. Rev. A 92, 022112 (2015)



% Non-Markovian Effects on the Dynamics of Entanglement 
\bibitem{bellomo2007non} B. Bellomo, R. L. Franco, G. Compagno, Phys.  Rev.  Lett.  \textbf{99}, 160502 (2007);

% Photonic State Tomography
\bibitem{tomography} J. B. Altepeter, E. R. Jeffrey, P. G. Kwiat, Advances in Atomic, Molecular, and Opt. Phys. {\bf 52}, 105 (2005). 


\end{thebibliography}
\end{document}